\documentclass{svproc}
\usepackage{url}
\usepackage{amsmath, amsfonts, amssymb}
\usepackage{mathtools}
\usepackage{hyperref}

\usepackage[ruled,vlined,lined,boxed,commentsnumbered]{algorithm2e}
\SetAlgorithmName{Algorithm}{algorithm}{List of Algorithms}
\SetKwInOut{Input}{Input}
\SetKwInOut{Output}{Output}

\usepackage{multirow}

\def\x{\mbox{\boldmath $x$}}
\def\bmu{\mbox{\boldmath $\mu$}}
\def\bsigma{\mbox{\boldmath $\Sigma$}}
\def\Real{\mbox{$\mathbb{R}$}}

\begin{document}
\mainmatter              % start of a contribution
\title{Portfolio Optimization of Indonesian Banking Stocks Using Robust Optimization}
\titlerunning{Portfolio Optimization}

\author{Visca Tri Winarty \and Sena Safarina}
\authorrunning{Visca Tri et al.}

\institute{
Sepuluh Nopember Institute of Technology, Surabaya, Indonesia\\
\email{safarina@its.ac.id}\\
\url{https://scholar.its.ac.id/en/persons/sena-safarina}
}

\maketitle
            % typeset the title of the contribution

\begin{abstract}
Since the COVID-19 pandemic, the number of investors in the Indonesia Stock Exchange has steadily increased, emphasizing the importance of portfolio optimization in balancing risk and return. The classical mean–variance optimization model, while widely applied, depends on historical return and risk estimates that are uncertain and may result in suboptimal portfolios. To address this limitation, robust optimization incorporates uncertainty sets to improve portfolio reliability under market fluctuations. This study constructs such sets using moving-window and bootstrapping methods and applies them to Indonesian banking stock data with varying risk-aversion parameters. The results show that robust optimization with the moving-window method, particularly with a smaller risk-aversion parameter, provides a better risk–return trade-off compared to the bootstrapping approach. These findings highlight the potential of the moving-window method to generate more effective portfolio strategies for risk-tolerant investors.

% We would like to encourage you to list your keywords within
% the abstract section using the \keywords{...} command.
\end{abstract}
\section{Introduction}
Since the emergence of the COVID-19 pandemic, the Indonesia Stock Exchange (IDX) has reported a gradual increase in the number of capital market investors. As of December 2022, the total number of investors had reached 10.23 million, representing an increase compared to previous years. This upward trend is projected to continue in the subsequent year~\cite{mulyana2022}. With the rising number of investors in Indonesia's capital market, the implementation of portfolio optimization becomes increasingly relevant for achieving balanced investment outcomes in terms of the risk-return trade-off.

Portfolio optimization is a fundamental concept in investment, with the primary objective of minimizing risk or maximizing expected returns. Harry Markowitz~\cite{markowitz1952} introduced the modern portfolio optimization model, which has been widely applied to guide investors in constructing an efficient portfolio. The classical model is formulated below.
\begin{align}
    \begin{array}{rcl}
    \min & : & \frac{\gamma}{2} \x^T \bsigma \x-\bmu^T \x, \\
    \text{s.t} & : & \pmb{e}^T \x = 1, \\
     & & 0 \le {x_i}\le 1 \quad \mbox{for} \quad i = 1, 2, \ldots, n.
    \end{array}\label{eq:MVO}
\end{align}

In model~\eqref{eq:MVO}, the important variable is $\pmb{x}\in\Real^n$, which denotes the vector of $n$-asset allocation. The objective is to minimize the trade-off between the risk, $\frac{\gamma}{2}\mbox{Var}(\pmb{r}^T\x)=\frac{\gamma}{2} \x^T \bsigma \x$, and the expected portfolio return, $E(\pmb{r}^T\x)=\bmu^T \x$. The constraint $\pmb{e}^T \x = 1$ ensures that the total asset allocation
equals $100\%$, with $\pmb{e}\in\Real^n$ denoting the vector of ones. Finally, the bound $0\leq x_i\leq 1$ for $i=1,2,\ldots,n$ indicates no short selling in this classical model.

In the objective function, $\gamma\in\Real_{++}$ represents the risk-aversion parameter and $\bmu^T=[\mu_1 \ \mu_2 \ \cdots \mu_n]$ denotes the vector of expected assets' return, computed by
\begin{equation}
    \mu_i=\frac{1}{m}\sum_{t=1}^m r_{i,t}\label{eq:mean}
\end{equation}
where $r_{i,t}$ is the daily return of asset-$i$ and for period-$t$ and $m$ is the length of observation period. Furthermore, the variance-covariance matrix $\bsigma$ can be obtained by
\begin{equation}
    \sigma_{i,j}=\frac{1}{m}\sum_{t=1}^m(r_{i,t}-\mu_i)(r_{j,t}-\mu_j)\label{eq:sigma}
\end{equation}
for $i,j\in\{1,2,\cdots,n\}$. 

The classical model~\eqref{eq:MVO} presumes that both parameters~\eqref{eq:mean} and~\eqref{eq:sigma} are known with certainty. However, in practice, the problem faces the uncertainty of future data, since both asset returns and risk measures are unknown and must be estimated from historical data~\cite{palomar2025portfolio}. However, these estimates are often subject to significant errors, as past observations may not accurately reflect future market conditions. To address such a limitation, the mean–variance optimization model is often reformulated as a robust optimization model~\cite{abdurakhman2022,costa2020,isavnin2019,tutuncu2004}. The key advantage of robust optimization lies in its ability to generate portfolio solutions that remain effective in worst-case scenarios. This approach has been widely applied in finance, particularly in the construction of optimal portfolios.

Tütüncü and Koenig~\cite{tutuncu2004} showed that robust optimization improves portfolio performance when uncertain inputs are represented within an uncertainty set. Several approaches can be used to construct such a set; for instance, Tütüncü and Koenig~\cite{tutuncu2004} proposed a moving-window method, while Abdurakhman~\cite{abdurakhman2022} employed bootstrapping. Motivated by these studies, this research adopts both methods to build the uncertainty set and compares their effects on the optimal portfolio solution obtained from the robust optimization model. The model is then solved in MATLAB for different risk-aversion parameters, where a smaller value reflects a higher tolerance for investment risk. In particular, this research focuses on implementing and comparing robust optimization with moving-window and bootstrapping methods using Indonesian banking stocks data.

The rest of this paper is ordered as follows. In Section 2, we explain robust portfolio optimization, as well as the implementation of moving window and bootstrapping methods to obtain the uncertainty set. Section 3 shows the numerical results for comparing robust portfolio optimization with moving window and bootstrapping methods. We also compare both results with the classical MVO. In Section 4, we present conclusions and discussions for future directions.

\section{Robust Optimization}
Robust optimization is an approach to solving optimization problems that require information about uncertainty sets in order to determine the worst-case scenarios to be optimized \cite{ben-tal1998}. Robust optimization modeling addresses the limitations of mean-variance optimization by formulating the problem based on the worst-case scenarios within the mean-variance optimization model \cite{cornuejols2006}. Problem~\eqref{eq:MVO} can be therefore reformulated using robust optimization as follows:
\begin{equation}\label{eq:robust}
\begin{array}{rcl}
\max & : & \left( \min\limits_{\boldsymbol{\mu}\in U_{\mu}} {\bmu}^T \x \right) 
          - \left( \max\limits_{\boldsymbol{\Sigma} \in U_{\Sigma}} \tfrac{1}{2}\gamma \, \x^T \bsigma \mathbf{x} \right) \\
\text{s.t.} & : & \boldsymbol{e}^T \boldsymbol{x} = 1,  \\
            &   & 0 \leq x_i \leq 1, \quad i = 1,\ldots,n.
\end{array}
\end{equation}
where, $U_{\mu}$ and $U_{\Sigma}$ denote uncertainty sets of the mean vector and the covariance matrix.

Various forms of uncertainty sets have been proposed in the literature. In this study, the uncertainty set is defined in the form of intervals, as introduced by \cite{engels2004}:
\begin{equation}
\begin{aligned}
U_{\mu} &= \left\{ \boldsymbol{\mu} \in \mathbb{R}^n \; \middle| \;
    \mu_i^0 - \beta_i \leq \mu_i \leq \mu_i^0 + \beta_i,\;
    i = 1,\ldots,n,\; \beta_i \geq 0 \right\}, \\
U_{\Sigma} &= \left\{ \boldsymbol{\Sigma} \in \mathbb{R}^{n \times n} \; \middle| \;
    \sigma_{ik}^0 - \delta_{ik} \leq \sigma_{ik} \leq \sigma_{ik}^0 + \delta_{ik},\;
    i,k = 1,\ldots,n,\; \delta_{ik} \geq 0 \right\}.
\end{aligned}\label{eq:uncertanty}
\end{equation}
These sets are constructed based on the interval bounds for the parameters:
\begin{align}
    \mu_i^l \leq \mu_i \leq \mu_i^u \quad \text{and} \quad
    \sigma_{ik}^l \leq \sigma_{ik} \leq \sigma_{ik}^u \nonumber
\end{align}
for $i,k\in\{1,2,\cdots,n\}$. Here, $\mu^l_i$ and $\mu^u_i$ represent lower and upper bounds of the asset return means, while $\sigma^l_{ik}$ and $\sigma^u_{ik}$ denote lower and upper bounds of the asset return covariances. The notation used in~\eqref{eq:uncertanty}, including $\mu_i^0$, $\sigma_{ik}^0$, $\beta_i$, and $\delta_{ik}$, is defined below:
\begin{align}
    \mu_i^0 = \frac{\mu_i^u + \mu_i^l}{2}, \ \beta_i = \frac{\mu_i^u - \mu_i^l}{2}, \ \sigma_{ik}^0 = \frac{\sigma_i^u + \sigma_i^l}{2} ,\ \text{and} \ \delta_{ik} = \frac{\sigma_i^u + \sigma_i^l}{2} \label{eq:allparam}
\end{align}

Thus, problem~\eqref{eq:robust} can be rewritten as
\begin{align}
    \begin{array}{rcl}
    \label{PersamaanRObust}
    \max &:& (\pmb{\mu^0})^T\pmb{x} -\pmb{\beta}^T\pmb{x} - \frac{1}{2}\gamma \pmb{x}^T \pmb{\Sigma^0} \pmb{x} - \frac{1}{2}\gamma \pmb{x}^T \pmb{\Delta} \pmb{x} \\
    \text{s.t.} &:& \pmb{e}^T\pmb{x} =1,  \\
     & & 0 \le {x_i}\le 1, \quad i=1,\ldots,n
    \end{array}
\end{align}
where
\begin{equation*}
    \pmb{\mu^0} = (\mu^0_1,\ldots,\mu^0_n)^T,  
    \pmb{\beta}=(\beta_1,\ldots,\beta_n)^T, 
    \pmb{\Sigma^0} = \begin{bmatrix}
        \sigma^0_{11} & \cdots & \sigma^0_{1n}\\
        \vdots & \ddots & \vdots\\
        \sigma^0_{n1} & \cdots & \sigma^0_{nn}
    \end{bmatrix},\text{and} \
    \pmb{\Delta} = \begin{bmatrix}
        \delta_{11} & \cdots & \delta_{1n}\\
        \vdots & \ddots & \vdots\\
        \delta_{n1} & \cdots & \delta_{nn}
    \end{bmatrix}.
\end{equation*}

 As mentioned previously, we will derive the uncertainty sets using two different approaches: the moving-window and bootstrapping.

\subsection{Moving-window Method}
The moving-window is a method for generating an uncertainty set by selecting a specific-sized window (a sub-sample of the data) to calculate the mean vector and covariance matrix for each window. The algorithm for constructing the uncertainty set is presented below:

\begin{algorithm}[H]
\caption{Construction of the Uncertainty Set using the Moving-Window Method \cite{tutuncu2004}}
\label{alg:movingwindow}
\Input{Number of assets $n$, number of time periods $m$}
\Output{Uncertainty set for the mean vector and covariance matrix}
\BlankLine
\textbf{Step 1:} Let $R \in \mathbb{R}^{m \times n}$ be the return matrix of $n$ assets over $m$ periods.\\
\textbf{Step 2:} Choose a window size $K$ such that $K < m$.\\
\textbf{Step 3:} For $t = 1$ to $m - K + 1$ \textbf{do}:\\
\Indp
    Extract the submatrix $R_t = [r_t, r_{t+1}, \ldots, r_{t+K-1}]$.\\
    Compute the mean vector $\boldsymbol{\mu}_t$ and covariance matrix $\boldsymbol{\Sigma}_t$ for $R_t$.\\
\Indm
\textbf{Step 4:} After all windows are processed:\\
\Indp
    For each asset $i$: 
    \begin{itemize}
        \item Compute $\mu_i^L = \min_t \mu_{t,i}$ and $\mu_i^U = \max_t \mu_{t,i}$.
    \end{itemize}
    For each covariance element $(i, k)$: 
    \begin{itemize}
        \item Compute $\sigma_{ik}^L = \min_t \sigma_{t,ik}$ and $\sigma_{ik}^U = \max_t \sigma_{t,ik}$.
    \end{itemize}
\Indm
\Return{Uncertainty set: $[\mu_i^L, \mu_i^U]$ and $[\sigma_{ik}^L, \sigma_{ik}^U]$ for all $i, k$}
\end{algorithm}

\subsection{Bootstrapping Method}
Unlike the moving-window method, the bootstrapping method utilizes random samples from the data population to generate an uncertainty set. This approach involves resampling the data, creating new samples from the existing dataset, and computing the mean vector and covariance matrix for each sample. The resulting uncertainty set is then used as the input for robust optimization model~\cite{abdurakhman2022}. The algorithm for constructing the uncertainty set using bootstrapping is presented below. In this algorithm, $\alpha$ is a confidence level where \cite{tutuncu2004} typically sets $\alpha=0.05$; and $N_{boot}$ is the number of bootstrap replications, which is usually chosen to be a large number.

\begin{algorithm}[H]
\caption{Construction of Uncertainty Set using Bootstrapping}
\label{alg:bootstrapping}
\Input{Number of assets $n$, number of time periods $m$, significance level $\alpha$}
\Output{Uncertainty set for mean vector $\boldsymbol{\mu}$ and covariance matrix $\boldsymbol{\Sigma}$}
\BlankLine
\textbf{Step 1:} Let $R \in \mathbb{R}^{m \times n}$ be the return matrix of $n$ assets over $m$ periods.\\
\textbf{Step 2:} Set block length $L = \lfloor m^{1/3} \rfloor$.\\
\textbf{Step 3:} Compute the number of blocks as $B = \lfloor m / L \rfloor$.\\
\textbf{Step 4:} \For{$i = 1$ \KwTo $N_{\text{boot}}$}{
    \For{$j = 1$ \KwTo $B$}{
        Randomly resample $L$ returns with replacement from $R$ to form block $R_{ij}$.\\
    }
    Concatenate the $B$ blocks to form bootstrap sample $R^{(i)}$.\\
    Compute sample mean $\mu^{(i)}$ and covariance matrix $\Sigma^{(i)}$ from $R^{(i)}$.\\
}
\textbf{Step 5:} For each asset $k$, define the uncertainty bounds as:
\begin{itemize}
    \item $\mu_k^l = \text{percentile}_{\alpha/2}\left(\mu_k^{(1)}, \ldots, \mu_k^{(N_{\text{boot}})}\right)$
    \item $\mu_k^u = \text{percentile}_{1-\alpha/2}\left(\mu_k^{(1)}, \ldots, \mu_k^{(N_{\text{boot}})}\right)$
\end{itemize}
\textbf{Step 6:} For each covariance entry $(p,q)$, define:
\begin{itemize}
    \item $\sigma_{pq}^l = \text{percentile}_{\alpha/2}\left(\sigma_{pq}^{(1)}, \ldots, \sigma_{pq}^{(N_{\text{boot}})}\right)$
    \item $\sigma_{pq}^u = \text{percentile}_{1-\alpha/2}\left(\sigma_{pq}^{(1)}, \ldots, \sigma_{pq}^{(N_{\text{boot}})}\right)$
\end{itemize}
\Return{Uncertainty set: $[\mu_k^l, \mu_k^u]$ and $[\sigma_{pq}^l, \sigma_{pq}^u]$ for all $k, p, q$}
\end{algorithm}

\section{Numerical Results}

We conducted numerical simulations of the robust optimization problem (\ref{PersamaanRObust}). The method was implemented using \texttt{Matlab R2021b} with the \texttt{CPLEX 12.10.0} setting and executed on a Windows 10 PC. The data set, taken from the \textit{Yahoo Finance}, consists of the daily closing stock of 45 banking companies in Indonesia during the period from March 25, 2022, to March 24, 2023, which spans 247 consecutive trading days. The stock codes of the banks used in this study are listed in Table~\ref{tab:bankcodes}.

Using the data, we first generated important parameters in portfolio optimization, return asset vector (\bmu) and variance-covariance matrix (\bsigma), using equation~\eqref{eq:mean} and~\eqref{eq:sigma}, respectively, as in~\eqref{eq:nilaimu} and~\eqref{eq:nilaisigma}. These two parameters are then used to construct the intervals that capture the uncertainty by applying moving-window and bootstrapping methods, following Algorithms~\ref{alg:movingwindow} and~\ref{alg:bootstrapping}, respectively.
\begin{equation}
    \bmu^T=\begin{bmatrix}-0.00439 & -0.00179 & -0.00055 & \cdots & 0.00306 & 0.00026 & 0.00023\end{bmatrix}\label{eq:nilaimu}
\end{equation}

\begin{equation}
\bsigma =
\begin{bmatrix}
0.00137 & 0.00026 & 0.00005 & \cdots & 0.00031 & 0.00021 & 0.00004 \\
0.00026 & 0.00048 & 0.00006 & \cdots & 0.00012 & 0.00015 & 0.00002 \\
0.00005 & 0.00006 & 0.00097 & \cdots & 0.00007 & -0.00004 & 0.00002 \\
\vdots  & \vdots  & \vdots  & \ddots & \vdots  & \vdots   & \vdots  \\
0.00031 & 0.00012 & 0.00007 & \cdots & 0.00182 & 0.00081 & 0.00003 \\
0.00021 & 0.00015 & -0.00004& \cdots & 0.00081 & 0.00124 & 0.00001 \\
0.00004 & 0.00002 & 0.00002 & \cdots & 0.00003 & 0.00001 & 0.00011 \\
\end{bmatrix}\label{eq:nilaisigma}
\end{equation}

\begin{table}
\centering
\caption{Stock codes of 45 Indonesian banking companies used in the dataset}
\label{tab:bankcodes}
\begin{tabular}{|c|c|c|c|c|c|c|c|c|c|}
\hline
$i$ & Asset Code & $i$ & Asset Code & $i$ & Asset Code & $i$ & Asset Code & $i$ & Asset Code \\ \hline
1  & AGRO & 10 & BBKP & 19 & BGTG & 28 & BNII & 37 & MASB \\
2  & AGRS & 11 & BBMD & 20 & BINA & 29 & BNLI & 38 & MAYA \\
3  & AMAR & 12 & BBNI & 21 & BJBR & 30 & BRIS & 39 & MCOR \\
4  & ARTO & 13 & BBRI & 22 & BJTM & 31 & BSIM & 40 & MEGA \\
5  & BABP & 14 & BBSI & 23 & BKSW & 32 & BTPN & 41 & NISP \\
6  & BACA & 15 & BBTN & 24 & BMAS & 33 & BTPS & 42 & NOBU \\
7  & BANK & 16 & BBYB & 25 & BMRI & 34 & BVIC & 43 & PNBN \\
8  & BBCA & 17 & BCIC & 26 & BNBA & 35 & DNAR & 44 & PNBS \\
9  & BBHI & 18 & BDMN & 27 & BNGA & 36 & INPC & 45 & SDRA \\
\hline
\end{tabular}
\end{table}

In the moving-window method, the uncertainty set is constructed by selecting a data sub-sample of fixed size. In this research, a window length of 90 days ($K=90$) is used, corresponding to quarterly financial reporting periods (February, May, August, and November), which are considered appropriate times for stock trading. With the chosen window length of $K=90$, the procedure requires 158 iterations to obtain the corresponding intervals.

Unlike the moving-window method, the length of bootstrap block $L$ is determined based on the sample size $m$. For 247 consecutive trading days, we have $m=247$, and based on \textbf{Step 2} in Algorithm~\ref{alg:bootstrapping},
\begin{equation*}
    L=247^{1/3}=6.2743.
\end{equation*}
Using this block length, the number of blocks required to form one bootstrap resample is
\begin{equation*}
    \frac{m}{L}=\frac{247}{6.2743}\approx 40.    
\end{equation*}
Then, both parameters $L$ and $\frac{m}{L}$ are used to construct the intervals in the bootstrap method. In addition, we set $N_{boot}=1000$ and $\alpha=0.05$.

The lower and upper bounds from the generated interval for the mean return vectors are presented in Table~\ref{table:IntervalMR}. The first column of the table indicates the method used, while the remaining columns show the lower bound $\mu_i^l$ and the upper bound $\mu_i^u$ for each asset code.
\begin{table}[h!]
\centering
\caption{Lower and Upper Bounds of the Mean Return from Two Methods.}
\label{tab:MiuCombined}
\begin{tabular}{|c|c|r|r|r|r|r|r|r|}
\hline
\textbf{Method} & \textbf{Bounds} & \textbf{AGRO} & \textbf{AGRS} & \textbf{AMAR} & \pmb{$\cdots$} & \textbf{PNBN} & \textbf{PNBS} & \textbf{SDRA} \\ \hline
\multirow{2}{*}{Moving-window} 
 & {$(\mu_i^l)_{MW}$} & -0.00669 & -0.00400 & -0.00501 & $\cdots$ & -0.00684 & -0.00412 & -0.00073 \\ \cline{2-9}
 & {$(\mu_i^u)_{MW}$} & -0.00024 & 0.00065 & 0.00292  & $\cdots$ & 0.01344  & 0.00521  & 0.00102  \\ \hline
\multirow{2}{*}{Bootstrapping} 
 & {$(\mu_i^l)_{B}$} & -0.01519 & -0.00852 & -0.00948 & $\cdots$ & -0.00927 & -0.00965 & -0.00277 \\ \cline{2-9}
 & {$(\mu_i^u)_{B}$} & 0.00793  & 0.00579  & 0.00957  & $\cdots$ & 0.01699  & 0.01156  & 0.00377  \\ \hline
\end{tabular}\label{table:IntervalMR}
\end{table}

In order to get the parameters for Robust Optimization, the lower and upper bounds from the moving-window method in Table \ref{tab:MiuCombined} is used to compute the vector $\pmb{\mu^0}$ and $\pmb{\beta}$ using the formula $\mu_i^0 = \frac{\mu_i^u+\mu_i^l}{2}$ and $\beta = \frac{\mu_i^u-\mu_i^l}{2}$ so that
\begin{equation*}
    \pmb{\mu}_{MW}^0 = (-0.00346; \ldots; 0.00014)^T \ \text{and} \ \pmb{\beta}_{MW} = (0.00323; \ldots ;0.00087)^T.
\end{equation*}
This is also done to the lower and upper bounds from the bootstrapping method:
\begin{align*}
    \pmb{\mu}_{B}^0 = (-0.00386; \ldots; 0.00014)^T \ \text{and} \
    \pmb{\beta}_{B} = (0.01200; \ldots ;0.00087)^T.
\end{align*}
All bounds are then used to compute the matrices $\pmb{\Sigma^0}$ and $\pmb{\Delta}$ using equation (6), based on the values presented in Tables~\ref{tab:SigmaMW} and~\ref{tab:SigmaBS}, resulting in the following matrices:

\begin{align}
 \pmb{\Sigma}_{MW}^0 = \begin{bmatrix}
0.00160 & \ldots & 0.00004\\
\vdots & \ddots & \vdots\\
0.00004 & \ldots & 0.00010
\end{bmatrix}
\text{and} \
\pmb{\Delta}_{MW} = \begin{bmatrix}
0.00051 & \ldots & 0.00006\\
\vdots &  \ddots & \vdots\\
0.00006 & \ldots & 0.00004
\end{bmatrix}.
\end{align}
Similarly, from the bootstrapping method, we obtain
\begin{equation}
       \pmb{\Sigma}_{B}^0 = \begin{bmatrix}
0.00190 & \ldots & 0.00003\\
\vdots & \ddots & \vdots\\
0.00003 & \ldots & 0.00013
\end{bmatrix}
\text{and} \
    \pmb{\Delta}_{B} = \begin{bmatrix}
0.00148 & \ldots & 0.00012\\
\vdots &  \ddots & \vdots\\
0.00012 & \ldots & 0.00007
\end{bmatrix}.
\end{equation}

\begin{table}
\caption{Lower and Upper Bounds of Return Covariance from moving-window Method.}
\label{tab:SigmaMW}
\resizebox{\textwidth}{!}{
\begin{tabular}{|c|c|c|c|c|c|c|c|c|}
\hline
                               & \textbf{Bounds}                           & \textbf{AGRO} & \textbf{AGRS} & \textbf{AMAR} & $\pmb{\cdots}$ & \textbf{PNBN} & \textbf{PNBS} & \textbf{SDRA} \\ \hline
\multirow{2}{*}{\textbf{AGRO}} & ${\sigma_{ik}^l}$ & 0.00109       & 0.00009       & -0.00002      & $\cdots$ & 0.00007       & 0.00004       & -0.00002      \\ \cline{2-9} 
                               & ${\sigma_{ik}^u}$ & 0.00211       & 0.00037       & 0.00012       & $\cdots$ & 0.00049       & 0.00030       & 0.00010       \\ \hline
\multirow{2}{*}{\textbf{AGRS}} & ${\sigma_{ik}^l}$ & 0.00009       & 0.00027       & -0.00001      & $\cdots$ & 0.00005       & -0.00006      & -0.00003      \\ \cline{2-9} 
                               & ${\sigma_{ik}^u}$ & 0.00037       & 0.00072       & 0.00015       & $\cdots$ & 0.00018       & 0.00027       & 0.00007       \\ \hline
\multirow{2}{*}{\textbf{AMAR}} & ${\sigma_{ik}^l}$ & -0.00002      & -0.00001      & 0.00065       & $\cdots$ & -0.00013      & -0.00044      & -0.00004      \\ \cline{2-9} 
                               & ${\sigma_{ik}^u}$ & 0.00012       & 0.00015       & 0.00120       & $\cdots$ & 0.00026       & 0.00030       & 0.00005       \\ \hline
                               $\pmb{\vdots}$ &   $\pmb{\vdots}$                                     &   $\pmb{\vdots}$            &   $\pmb{\vdots}$            &   $\pmb{\vdots}$            & $\pmb{\ddots}$ &     $\pmb{\vdots}$           &   $\pmb{\vdots}$             &       $\pmb{\vdots}$         \\ \hline
\multirow{2}{*}{\textbf{PNBN}} & ${\sigma_{ik}^l}$ & 0.00007       & 0.00005       & -0.00013      & $\cdots$ & 0.00100       & 0.00033       & -0.00008      \\ \cline{2-9} 
                               & ${\sigma_{ik}^u}$ & 0.00049       & 0.00018       & 0.00026       & $\cdots$ & 0.00292       & 0.00137       & 0.00007       \\ \hline
\multirow{2}{*}{\textbf{PNBS}} & ${\sigma_{ik}^l}$ & 0.00004       & -0.00006      & -0.00044      & $\cdots$ & 0.00033       & 0.00045       & -0.00008      \\ \cline{2-9} 
                               & ${\sigma_{ik}^u}$ & 0.00030       & 0.00027       & 0.00030       & $\cdots$ & 0.00137       & 0.00223       & 0.00006       \\ \hline
\multirow{2}{*}{\textbf{SDRA}} & ${\sigma_{ik}^l}$ & -0.00002      & -0.00003      & -0.00004      & $\cdots$ & -0.00008      & -0.00008      & 0.00006       \\ \cline{2-9} 
                               & ${\sigma_{ik}^u}$ & 0.00010       & 0.00007       & 0.00005       & $\cdots$ & 0.00007       & 0.00006       & 0.00014       \\ \hline
\end{tabular}
}
\end{table} 

\begin{table}
\begin{center}
\caption{Lower and Upper Bounds of Return Covariance from Bootstrapping Method.}
\label{tab:SigmaBS}
\resizebox{\textwidth}{!}{
\begin{tabular}{|c|c|c|c|c|c|c|c|c|}
\hline
                               & \textbf{Bounds}                           & \textbf{AGRO} & \textbf{AGRS} & \textbf{AMAR} & $\pmb{\cdots}$ & \textbf{PNBN} & \textbf{PNBS} & \textbf{SDRA} \\ \hline
\multirow{2}{*}{\textbf{AGRO}} & $\sigma_{ik}^l$ & 0.00054       & 0.00004       & -0.00028      & $\cdots$ & -0.00017      & -0.00017      & -0.00010      \\ \cline{2-9} 
                               & $\sigma_{ik}^u$ & 0.00351       & 0.00059       & 0.00038       & $\cdots$ & 0.00081       & 0.00058       & 0.00014       \\ \hline
\multirow{2}{*}{\textbf{AGRS}} & $\sigma_{ik}^l$ & 0.00004       & 0.00023       & -0.00014      & $\cdots$& -0.00017      & -0.00011      & -0.00005      \\ \cline{2-9} 
                               & $\sigma_{ik}^u$ & 0.00059       & 0.00079       & 0.00026       & $\cdots$ & 0.00044       & 0.00056       & 0.00010       \\ \hline
\multirow{2}{*}{\textbf{AMAR}} & $\sigma_{ik}^l$ & -0.00028      & -0.00014      & 0.00050       & $\cdots$ & -0.00030      & -0.00062      & -0.00010      \\ \cline{2-9} 
                               & $\sigma_{ik}^u$ & 0.00038       & 0.00026       & 0.00159       & $\cdots$ & 0.00049       & 0.00035       & 0.00012       \\ \hline
                             $\pmb{\vdots}$  & $\vdots$                                         &  $\vdots$             &  $\vdots$             & $\vdots$              & $\ddots$ &  $\vdots$             &   $\vdots$            &  $\vdots$             \\ \hline
\multirow{2}{*}{\textbf{PNBN}} & $\sigma_{ik}^l$ & -0.00017      & -0.00017      & -0.00030      & $\cdots$ & 0.00074       & 0.00023       & -0.00014      \\ \cline{2-9} 
                               & $\sigma_{ik}^u$ & 0.00081       & 0.00044       & 0.00049       & $\cdots$ & 0.00344       & 0.00176       & 0.00018       \\ \hline
\multirow{2}{*}{\textbf{PNBS}} & $\sigma_{ik}^l$ & -0.00017      & -0.00011      & -0.00062      & $\cdots$ & 0.00023       & 0.00050       & -0.00016      \\ \cline{2-9} 
                               & $\sigma_{ik}^u$ & 0.00058       & 0.00056       & 0.00035       & $\cdots$ & 0.00176       & 0.00316       & 0.00012       \\ \hline
\multirow{2}{*}{\textbf{SDRA}} & $\sigma_{ik}^l$ & -0.00010      & -0.00005      & -0.00010      & $\cdots$ & -0.00014      & -0.00016      & 0.00006       \\ \cline{2-9} 
                               & $\sigma_{ik}^u$ & 0.00014       & 0.00010       & 0.00012       & $\cdots$ & 0.00018       & 0.00012       & 0.00019       \\ \hline
\end{tabular}
}
\end{center}
\end{table}
\noindent
Once the values of $\pmb{\mu}^0$, $\pmb{\beta}$, $\pmb{\Sigma}^0$ and $\pmb{\Delta}$ are obtained, they are substituted into~(\ref{PersamaanRObust}) to derive the robust optimization solution, yielding results such as selected assets, asset proportions, and the objective function value.

\begin{table}
\caption{Allocation of Chosen Asset's Proportion and Objective Function Value.}
\label{tab:proporsi}
\resizebox{\textwidth}{!}{
\begin{tabular}{|c|ccc|ccc|}
\hline
\multirow{2}{*}{\textbf{Proportion}} & \multicolumn{3}{c|}{\textit{\textbf{Moving   Window}}}                                           & \multicolumn{3}{c|}{\textit{\textbf{Bootstrapping}}}                                             \\ \cline{2-7} 
                                   & \multicolumn{1}{c|}{$\pmb{\gamma = 5}$}       & \multicolumn{1}{c|}{$\pmb{\gamma = 50}$}      & $\pmb{\gamma = 100}$     & \multicolumn{1}{c|}{$\pmb{\gamma = 5}$}       & \multicolumn{1}{c|}{$\pmb{\gamma = 50}$}      & $\pmb{\gamma = 100}$       \\ \hline
\textbf{BBCA}                      & \multicolumn{1}{c|}{0.07923}          & \multicolumn{1}{c|}{0.04409}          & 0.01827          & \multicolumn{1}{c|}{0}                & \multicolumn{1}{c|}{0}                & 0                \\ \hline
\textbf{BBMD}                      & \multicolumn{1}{c|}{0}                & \multicolumn{1}{c|}{0.04584}          & 0.05815          & \multicolumn{1}{c|}{0}                & \multicolumn{1}{c|}{0}                & 0.00994          \\ \hline
\textbf{BBNI}                      & \multicolumn{1}{c|}{0}                & \multicolumn{1}{c|}{0}                & 0                & \multicolumn{1}{c|}{0}                & \multicolumn{1}{c|}{0}                & 0                \\ \hline
\textbf{BDMN}                      & \multicolumn{1}{c|}{0}                & \multicolumn{1}{c|}{0}                & 0                & \multicolumn{1}{c|}{0}                & \multicolumn{1}{c|}{0}                & 0                \\ \hline
\textbf{BINA}                      & \multicolumn{1}{c|}{0.10324}          & \multicolumn{1}{c|}{0.03235}          & 0.02277          & \multicolumn{1}{c|}{0}                & \multicolumn{1}{c|}{0}                & 0                \\ \hline
\textbf{BJBR}                      & \multicolumn{1}{c|}{0}                & \multicolumn{1}{c|}{0.10305}          & 0.11073          & \multicolumn{1}{c|}{0}                & \multicolumn{1}{c|}{0.13762}          & 0.1508           \\ \hline
\textbf{BJTM}                      & \multicolumn{1}{c|}{0}                & \multicolumn{1}{c|}{0.19754}          & 0.24193          & \multicolumn{1}{c|}{0.44471}          & \multicolumn{1}{c|}{0.4168}           & 0.38734          \\ \hline
\textbf{BMAS}                      & \multicolumn{1}{c|}{0}                & \multicolumn{1}{c|}{0}                & 0                & \multicolumn{1}{c|}{0}                & \multicolumn{1}{c|}{0}                & 0                \\ \hline
\textbf{BMRI}                      & \multicolumn{1}{c|}{0.08743}          & \multicolumn{1}{c|}{0}                & 0                & \multicolumn{1}{c|}{0}                & \multicolumn{1}{c|}{0}                & 0                \\ \hline
\textbf{BNGA}                      & \multicolumn{1}{c|}{0.52335}          & \multicolumn{1}{c|}{0.14778}          & 0.1045           & \multicolumn{1}{c|}{0}                & \multicolumn{1}{c|}{0.03982}          & 0.03897          \\ \hline
\textbf{BNII}                      & \multicolumn{1}{c|}{0}                & \multicolumn{1}{c|}{0}                & 0.01915          & \multicolumn{1}{c|}{0}                & \multicolumn{1}{c|}{0}                & 0.02309          \\ \hline
\textbf{BNLI}                      & \multicolumn{1}{c|}{0}                & \multicolumn{1}{c|}{0}                & 0.01231          & \multicolumn{1}{c|}{0}                & \multicolumn{1}{c|}{0}                & 0.0038           \\ \hline
\textbf{BRIS}                      & \multicolumn{1}{c|}{0}                & \multicolumn{1}{c|}{0}                & 0                & \multicolumn{1}{c|}{0}                & \multicolumn{1}{c|}{0}                & 0                \\ \hline
\textbf{BSIM}                      & \multicolumn{1}{c|}{0}                & \multicolumn{1}{c|}{0.03671}          & 0.03587          & \multicolumn{1}{c|}{0}                & \multicolumn{1}{c|}{0.00971}          & 0.01612          \\ \hline
\textbf{BTPN}                      & \multicolumn{1}{c|}{0}                & \multicolumn{1}{c|}{0.00123}          & 0.02361          & \multicolumn{1}{c|}{0}                & \multicolumn{1}{c|}{0.04952}          & 0.05379          \\ \hline
\textbf{MASB}                      & \multicolumn{1}{c|}{0}                & \multicolumn{1}{c|}{0.06208}          & 0.05241          & \multicolumn{1}{c|}{0}                & \multicolumn{1}{c|}{0}                & 0.00391          \\ \hline
\textbf{MEGA}                      & \multicolumn{1}{c|}{0}                & \multicolumn{1}{c|}{0}                & 0.00459          & \multicolumn{1}{c|}{0}                & \multicolumn{1}{c|}{0}                & 0                \\ \hline
\textbf{NISP}                      & \multicolumn{1}{c|}{0.00482}          & \multicolumn{1}{c|}{0.07177}          & 0.06518          & \multicolumn{1}{c|}{0.02681}          & \multicolumn{1}{c|}{0.08367}          & 0.07791          \\ \hline
\textbf{PNBN}                      & \multicolumn{1}{c|}{0}                & \multicolumn{1}{c|}{0}                & 0                & \multicolumn{1}{c|}{0}                & \multicolumn{1}{c|}{0}                & 0                \\ \hline
\textbf{PNBS}                      & \multicolumn{1}{c|}{0}                & \multicolumn{1}{c|}{0}                & 0                & \multicolumn{1}{c|}{0}                & \multicolumn{1}{c|}{0}                & 0                \\ \hline
\textbf{SDRA}                      & \multicolumn{1}{c|}{0.20192}          & \multicolumn{1}{c|}{0.25755}          & 0.24126          & \multicolumn{1}{c|}{0.52848}          & \multicolumn{1}{c|}{0.26286}          & 0.23433          \\ \hline
\textbf{$f_{val}$}                      & \multicolumn{1}{c|}{\textbf{0.00076}} & \multicolumn{1}{c|}{\textbf{0.00221}} & \textbf{0.00286} & \multicolumn{1}{c|}{\textbf{0.00313}} & \multicolumn{1}{c|}{\textbf{0.00499}} & \textbf{0.00694} \\ \hline
\end{tabular}
}
\end{table}
Table \ref{tab:proporsi} shows the results of selecting different values of $\gamma$. The first column lists the assets, the second to fourth columns present the asset allocations obtained using uncertainty sets generated from the moving-window method, while the remaining columns present the allocations derived from the bootstrapping method.

 In this research, three levels of the risk aversion coefficient were considered, namely $\gamma = \{5, 50,$ and $100\}$. These values were selected to represent different degrees of risk preference, ranging from relatively low to high, and to observe how the robust optimization model responds to variations in risk aversion. Although these values are not standard benchmarks, they provide representative cases for exploratory analysis. A broader sensitivity analysis with additional $\gamma$ values is left for future research.
 
 It can be observed from the table that different values of $\gamma$ lead to different asset selections in constructing the portfolio. For instance, the moving-window method produces an optimal allocation of 7.92\% in stock BBCA, 10.32\% in stock BINA, 8.74\% in stock BMRI, 52.34\% in stock BNGA, 0.48\% in stock NISP, and 20.19\% in stock SDRA. In contrast, the bootstrapping method yields a different set of selected assets, with optimal allocations of 44.47\% in stock BJTM, 2.68\% in stock NISP, and 52.85\% in stock SDRA. Furthermore, the objective value $(f_{val})$, which reflects the risk–return trade-off, is lower under the moving-window method compared to the bootstrapping method, indicating a more favorable balance between risk and return.
 
Moreover, the objective function values confirm that, across different methods and risk aversion coefficients $\gamma$, robust optimization with uncertainty sets derived from the moving-window method consistently produces better results than those derived from the bootstrapping method, as shown by its smaller objective values. This can be explained by the fact that, although both methods use the same historical data, the moving-window method relies only on the most recent subset of data, making the resulting uncertainty sets more reflective of current market conditions, whereas the bootstrapping method resamples from the entire dataset and thus captures a wider range of variability, including past periods that may no longer represent the present market situation. Consequently, the moving-window method often yields narrower and more relevant uncertainty sets, while the bootstrapping method produces more conservative allocations. As explained earlier, we used 90-day window length ($K = 90$), corresponding to quarterly financial reporting periods (February, May, August, and November). This choice provides a reasonable balance between capturing recent market dynamics and avoiding overfitting to very short-term fluctuations. To further illustrate the practical implications of these optimization results, we present an investment scenario.

Suppose an investor intends to invest Rp. 100,000.00 (Indonesian Rupiah) on 24 March 2023. The optimal stock proportions obtained from Table \ref{tab:AlokasiMW} are multiplied by the investment capital of Rp. 100,000.00 to obtain the stock price allocation. Subsequently, the number of shares is obtained by dividing the allocated amount by the closing price on 24 March 2023. Tables \ref{tab:AlokasiMW} and \ref{tab:AlokasiBS} present the resulting investment allocation from the moving-window and bootstrapping methods, respectively.

\begin{table}
\caption{Fund Allocation for Portfolio from Robust Optimization with moving-window.}
\label{tab:AlokasiMW}
\resizebox{\textwidth}{!}{
\begin{tabular}{|c|c|lclclc|}
\hline
\multirow{3}{*}{\textbf{Asset}} & \multirow{3}{*}{\textbf{\begin{tabular}[c]{@{}c@{}}Price\\ (24/3/2023)\end{tabular}}} & \multicolumn{6}{c|}{\textit{\textbf{moving-window}}}                                                                                                                                                                  \\ \cline{3-8} 
                               &                                                                                       & \multicolumn{2}{c|}{$\pmb{\gamma = 5}$}                                                       & \multicolumn{2}{c|}{$\pmb{\gamma = 50}$}                                                      & \multicolumn{2}{c|}{$\pmb{\gamma = 100}$}                                \\ \cline{3-8} 
                               &                                                                                       & \multicolumn{1}{c|}{\textbf{Alokasi}} & \multicolumn{1}{c|}{\textbf{Lembar}} & \multicolumn{1}{c|}{\textbf{Alokasi}} & \multicolumn{1}{c|}{\textbf{Lembar}} & \multicolumn{1}{c|}{\textbf{Alokasi}} & \textbf{Lembar} \\ \hline
\textbf{BBCA}                  & 8825                                                                                  & \multicolumn{1}{l|}{Rp7,923.00}       & \multicolumn{1}{c|}{1}               & \multicolumn{1}{l|}{Rp4,409.00}       & \multicolumn{1}{c|}{0}               & \multicolumn{1}{l|}{Rp1,827.00}       & 0               \\ \hline
\textbf{BBMD}                  & 1960                                                                                  & \multicolumn{1}{l|}{Rp0.00}           & \multicolumn{1}{c|}{0}               & \multicolumn{1}{l|}{Rp4,584.00}       & \multicolumn{1}{c|}{2}               & \multicolumn{1}{l|}{Rp5,815.00}       & 3               \\ \hline
\textbf{BBNI}                  & 9625                                                                                  & \multicolumn{1}{l|}{Rp0.00}           & \multicolumn{1}{c|}{0}               & \multicolumn{1}{l|}{Rp0.00}           & \multicolumn{1}{c|}{0}               & \multicolumn{1}{l|}{Rp0.00}           & 0               \\ \hline
\textbf{BDMN}                  & 2830                                                                                  & \multicolumn{1}{l|}{Rp0.00}           & \multicolumn{1}{c|}{0}               & \multicolumn{1}{l|}{Rp0.00}           & \multicolumn{1}{c|}{0}               & \multicolumn{1}{l|}{Rp0.00}           & 0               \\ \hline
\textbf{BINA}                  & 3990                                                                                  & \multicolumn{1}{l|}{Rp10,324.00}      & \multicolumn{1}{c|}{3}               & \multicolumn{1}{l|}{Rp3,235.00}       & \multicolumn{1}{c|}{1}               & \multicolumn{1}{l|}{Rp2,277.00}       & 1               \\ \hline
\textbf{BJBR}                  & 1335                                                                                  & \multicolumn{1}{l|}{Rp0.00}           & \multicolumn{1}{c|}{0}               & \multicolumn{1}{l|}{Rp10,305.00}      & \multicolumn{1}{c|}{8}               & \multicolumn{1}{l|}{Rp11,073.00}      & 8               \\ \hline
\textbf{BJTM}                  & 735                                                                                   & \multicolumn{1}{l|}{Rp0.00}           & \multicolumn{1}{c|}{0}               & \multicolumn{1}{l|}{Rp19,754.00}      & \multicolumn{1}{c|}{27}              & \multicolumn{1}{l|}{Rp24,193.00}      & 33              \\ \hline
\textbf{BMAS}                  & 1395                                                                                  & \multicolumn{1}{l|}{Rp0.00}           & \multicolumn{1}{c|}{0}               & \multicolumn{1}{l|}{Rp0.00}           & \multicolumn{1}{c|}{0}               & \multicolumn{1}{l|}{Rp0.00}           & 0               \\ \hline
\textbf{BMRI}                  & 5450                                                                                  & \multicolumn{1}{l|}{Rp8,743.00}       & \multicolumn{1}{c|}{2}               & \multicolumn{1}{l|}{Rp0.00}           & \multicolumn{1}{c|}{0}               & \multicolumn{1}{l|}{Rp0.00}           & 0               \\ \hline
\textbf{BNGA}                  & 1225                                                                                  & \multicolumn{1}{l|}{Rp52,335.00}      & \multicolumn{1}{c|}{43}              & \multicolumn{1}{l|}{Rp14,778.00}      & \multicolumn{1}{c|}{12}              & \multicolumn{1}{l|}{Rp10,450.00}      & 9               \\ \hline
\textbf{BNII}                  & 226                                                                                   & \multicolumn{1}{l|}{Rp0.00}           & \multicolumn{1}{c|}{0}               & \multicolumn{1}{l|}{Rp0.00}           & \multicolumn{1}{c|}{0}               & \multicolumn{1}{l|}{Rp1,915.00}       & 8               \\ \hline
\textbf{BNLI}                  & 930                                                                                   & \multicolumn{1}{l|}{Rp0.00}           & \multicolumn{1}{c|}{0}               & \multicolumn{1}{l|}{Rp0.00}           & \multicolumn{1}{c|}{0}               & \multicolumn{1}{l|}{Rp1,231.00}       & 1               \\ \hline
\textbf{BRIS}                  & 1610                                                                                  & \multicolumn{1}{l|}{Rp0.00}           & \multicolumn{1}{c|}{0}               & \multicolumn{1}{l|}{Rp0.00}           & \multicolumn{1}{c|}{0}               & \multicolumn{1}{l|}{Rp0.00}           & 0               \\ \hline
\textbf{BSIM}                  & 890                                                                                   & \multicolumn{1}{l|}{Rp0.00}           & \multicolumn{1}{c|}{0}               & \multicolumn{1}{l|}{Rp3,671.00}       & \multicolumn{1}{c|}{4}               & \multicolumn{1}{l|}{Rp3,587.00}       & 4               \\ \hline
\textbf{BTPN}                  & 2490                                                                                  & \multicolumn{1}{l|}{Rp0.00}           & \multicolumn{1}{c|}{0}               & \multicolumn{1}{l|}{Rp123.00}         & \multicolumn{1}{c|}{0}               & \multicolumn{1}{l|}{Rp2,361.00}       & 1               \\ \hline
\textbf{MASB}                  & 3420                                                                                  & \multicolumn{1}{l|}{Rp0.00}           & \multicolumn{1}{c|}{0}               & \multicolumn{1}{l|}{Rp6,208.00}       & \multicolumn{1}{c|}{2}               & \multicolumn{1}{l|}{Rp5,241.00}       & 2               \\ \hline
\textbf{MEGA}                  & 5075                                                                                  & \multicolumn{1}{l|}{Rp0.00}           & \multicolumn{1}{c|}{0}               & \multicolumn{1}{l|}{Rp0.00}           & \multicolumn{1}{c|}{0}               & \multicolumn{1}{l|}{Rp459.00}         & 0               \\ \hline
\textbf{NISP}                  & 755                                                                                   & \multicolumn{1}{l|}{Rp482.00}         & \multicolumn{1}{c|}{1}               & \multicolumn{1}{l|}{Rp7,177.00}       & \multicolumn{1}{c|}{10}              & \multicolumn{1}{l|}{Rp6,518.00}       & 9               \\ \hline
\textbf{PNBN}                  & 1415                                                                                  & \multicolumn{1}{l|}{Rp0.00}           & \multicolumn{1}{c|}{0}               & \multicolumn{1}{l|}{Rp0.00}           & \multicolumn{1}{c|}{0}               & \multicolumn{1}{l|}{Rp0.00}           & 0               \\ \hline
\textbf{PNBS}                  & 58                                                                                    & \multicolumn{1}{l|}{Rp0.00}           & \multicolumn{1}{c|}{0}               & \multicolumn{1}{l|}{Rp0.00}           & \multicolumn{1}{c|}{0}               & \multicolumn{1}{l|}{Rp0.00}           & 0               \\ \hline
\textbf{SDRA}                  & 585                                                                                   & \multicolumn{1}{l|}{Rp20,192.00}      & \multicolumn{1}{c|}{35}              & \multicolumn{1}{l|}{Rp25,755.00}      & \multicolumn{1}{c|}{44}              & \multicolumn{1}{l|}{Rp23,054.00}      & 39              \\ \hline
\end{tabular}
}
\end{table}
\begin{table}
\caption{Fund Allocation for Portfolio from Robust Optimization with Bootstrapping.}
\label{tab:AlokasiBS}
\resizebox{\textwidth}{!}{
\begin{tabular}{|c|c|cccccc|}
\hline
\multirow{3}{*}{\textbf{Asset}} & \multirow{3}{*}{\textbf{\begin{tabular}[c]{@{}c@{}}Price\\ (24/3/2023)\end{tabular}}} & \multicolumn{6}{c|}{\textit{\textbf{Bootstrapping}}}                                                                                                                                                    \\ \cline{3-8} 
                               &                                                                                       & \multicolumn{2}{c|}{$\pmb{\gamma = 5}$}                                                       & \multicolumn{2}{c|}{$\pmb{\gamma = 50}$}                                                      & \multicolumn{2}{c|}{$\pmb{\gamma = 100}$}                  \\ \cline{3-8} 
                               &                                                                                       & \multicolumn{1}{c|}{\textbf{Alokasi}} & \multicolumn{1}{c|}{\textbf{Lembar}} & \multicolumn{1}{c|}{\textbf{Alokasi}} & \multicolumn{1}{c|}{\textbf{Lembar}} & \multicolumn{1}{c|}{Alokasi}     & Lembar \\ \hline
\textbf{BBCA}                  & 8825                                                                                  & \multicolumn{1}{c|}{Rp0.00}           & \multicolumn{1}{c|}{0}               & \multicolumn{1}{c|}{Rp0.00}           & \multicolumn{1}{c|}{0}               & \multicolumn{1}{c|}{Rp0.00}      & 0      \\ \hline
\textbf{BBMD}                  & 1960                                                                                  & \multicolumn{1}{c|}{Rp0.00}           & \multicolumn{1}{c|}{0}               & \multicolumn{1}{c|}{Rp0.00}           & \multicolumn{1}{c|}{0}               & \multicolumn{1}{c|}{Rp994.00}    & 1      \\ \hline
\textbf{BBNI}                  & 9625                                                                                  & \multicolumn{1}{c|}{Rp0.00}           & \multicolumn{1}{c|}{0}               & \multicolumn{1}{c|}{Rp0.00}           & \multicolumn{1}{c|}{0}               & \multicolumn{1}{c|}{Rp0.00}      & 0      \\ \hline
\textbf{BDMN}                  & 2830                                                                                  & \multicolumn{1}{c|}{Rp0.00}           & \multicolumn{1}{c|}{0}               & \multicolumn{1}{c|}{Rp0.00}           & \multicolumn{1}{c|}{0}               & \multicolumn{1}{c|}{Rp0.00}      & 0      \\ \hline
\textbf{BINA}                  & 3990                                                                                  & \multicolumn{1}{c|}{Rp0.00}           & \multicolumn{1}{c|}{0}               & \multicolumn{1}{c|}{Rp0.00}           & \multicolumn{1}{c|}{0}               & \multicolumn{1}{c|}{Rp0.00}      & 0      \\ \hline
\textbf{BJBR}                  & 1335                                                                                  & \multicolumn{1}{c|}{Rp0.00}           & \multicolumn{1}{c|}{0}               & \multicolumn{1}{c|}{Rp13,762.00}      & \multicolumn{1}{c|}{10}              & \multicolumn{1}{c|}{Rp15,080.00} & 11     \\ \hline
\textbf{BJTM}                  & 735                                                                                   & \multicolumn{1}{c|}{Rp44,471.00}      & \multicolumn{1}{c|}{61}              & \multicolumn{1}{c|}{Rp41,680.00}      & \multicolumn{1}{c|}{57}              & \multicolumn{1}{c|}{Rp38,734.00} & 53     \\ \hline
\textbf{BMAS}                  & 1395                                                                                  & \multicolumn{1}{c|}{Rp0.00}           & \multicolumn{1}{c|}{0}               & \multicolumn{1}{c|}{Rp0.00}           & \multicolumn{1}{c|}{0}               & \multicolumn{1}{c|}{Rp0.00}      & 0      \\ \hline
\textbf{BMRI}                  & 5450                                                                                  & \multicolumn{1}{c|}{Rp0.00}           & \multicolumn{1}{c|}{0}               & \multicolumn{1}{c|}{Rp0.00}           & \multicolumn{1}{c|}{0}               & \multicolumn{1}{c|}{Rp0.00}      & 0      \\ \hline
\textbf{BNGA}                  & 1225                                                                                  & \multicolumn{1}{c|}{Rp0.00}           & \multicolumn{1}{c|}{0}               & \multicolumn{1}{c|}{Rp3,982.00}       & \multicolumn{1}{c|}{3}               & \multicolumn{1}{c|}{Rp3,897.00}  & 3      \\ \hline
\textbf{BNII}                  & 226                                                                                   & \multicolumn{1}{c|}{Rp0.00}           & \multicolumn{1}{c|}{0}               & \multicolumn{1}{c|}{Rp0.00}           & \multicolumn{1}{c|}{0}               & \multicolumn{1}{c|}{Rp2,309.00}  & 10     \\ \hline
\textbf{BNLI}                  & 930                                                                                   & \multicolumn{1}{c|}{Rp0.00}           & \multicolumn{1}{c|}{0}               & \multicolumn{1}{c|}{Rp0.00}           & \multicolumn{1}{c|}{0}               & \multicolumn{1}{c|}{Rp380.00}    & 0      \\ \hline
\textbf{BRIS}                  & 1610                                                                                  & \multicolumn{1}{c|}{Rp0.00}           & \multicolumn{1}{c|}{0}               & \multicolumn{1}{c|}{Rp0.00}           & \multicolumn{1}{c|}{0}               & \multicolumn{1}{c|}{Rp0.00}      & 0      \\ \hline
\textbf{BSIM}                  & 890                                                                                   & \multicolumn{1}{c|}{Rp0.00}           & \multicolumn{1}{c|}{0}               & \multicolumn{1}{c|}{Rp971.00}         & \multicolumn{1}{c|}{1}               & \multicolumn{1}{c|}{Rp1,612.00}  & 2      \\ \hline
\textbf{BTPN}                  & 2490                                                                                  & \multicolumn{1}{c|}{Rp0.00}           & \multicolumn{1}{c|}{0}               & \multicolumn{1}{c|}{Rp4,952.00}       & \multicolumn{1}{c|}{2}               & \multicolumn{1}{c|}{Rp5,379.00}  & 2      \\ \hline
\textbf{MASB}                  & 3420                                                                                  & \multicolumn{1}{c|}{Rp0.00}           & \multicolumn{1}{c|}{0}               & \multicolumn{1}{c|}{Rp0.00}           & \multicolumn{1}{c|}{0}               & \multicolumn{1}{c|}{Rp391.00}    & 0      \\ \hline
\textbf{MEGA}                  & 5075                                                                                  & \multicolumn{1}{c|}{Rp0.00}           & \multicolumn{1}{c|}{0}               & \multicolumn{1}{c|}{Rp0.00}           & \multicolumn{1}{c|}{0}               & \multicolumn{1}{c|}{Rp0.00}      & 0      \\ \hline
\textbf{NISP}                  & 755                                                                                   & \multicolumn{1}{c|}{Rp2,681.00}       & \multicolumn{1}{c|}{4}               & \multicolumn{1}{c|}{Rp8,367.00}       & \multicolumn{1}{c|}{11}              & \multicolumn{1}{c|}{Rp7,791.00}  & 10     \\ \hline
\textbf{PNBN}                  & 1415                                                                                  & \multicolumn{1}{c|}{Rp0.00}           & \multicolumn{1}{c|}{0}               & \multicolumn{1}{c|}{Rp0.00}           & \multicolumn{1}{c|}{0}               & \multicolumn{1}{c|}{Rp0.00}      & 0      \\ \hline
\textbf{PNBS}                  & 58                                                                                    & \multicolumn{1}{c|}{Rp0.00}           & \multicolumn{1}{c|}{0}               & \multicolumn{1}{c|}{Rp0.00}           & \multicolumn{1}{c|}{0}               & \multicolumn{1}{c|}{Rp0.00}      & 0      \\ \hline
\textbf{SDRA}                  & 585                                                                                   & \multicolumn{1}{c|}{Rp52,848.00}      & \multicolumn{1}{c|}{90}              & \multicolumn{1}{c|}{Rp26,286.00}      & \multicolumn{1}{c|}{45}              & \multicolumn{1}{c|}{Rp23,433.00} & 40     \\ \hline
\end{tabular}
}
\end{table}

The performance measure used to evaluate portfolio outcome is capital gain, which refers to the positive difference or profit obtained when selling an asset at a price higher than the purchase price. Conversely, a capital loss occurs when the asset is sold at a lower price \cite{hermuningsih2019}. Thus, profit or loss is calculated by subtracting the purchase price from the selling price and multiplying it by the number of shares. Table \ref{tab:pasarbagus} presents the profit or loss obtained under favorable stock market conditions for each optimization method. For instance, if an investor wants to invest on 30 March 2023 in a good market condition, the moving-window method yields profits of Rp. 11,715.00 for $\gamma = 5$, Rp. 1,030.00 for $\gamma = 50$, and Rp. 892.00 for $\gamma = 100$. Similarly, the bootstrapping method yields profits of Rp. 60.00 for $\gamma = 5$, Rp. 625.00 for $\gamma = 50$, and Rp. 685.00 for $\gamma = 100$.

\begin{table}
\caption{Capital Gain (Loss) on Good Market Condition.}
\label{tab:pasarbagus}
\resizebox{\textwidth}{!}{
\begin{tabular}{|cc|lll|lll|}
\hline
\multicolumn{1}{|c|}{\multirow{2}{*}{\textbf{Asset}}} & \multirow{2}{*}{\textbf{\begin{tabular}[c]{@{}c@{}}Price\\  (30/3/2023)\end{tabular}}} & \multicolumn{3}{c|}{\textit{\textbf{Moving   Window}}}                                        & \multicolumn{3}{c|}{\textit{\textbf{Bootstrapping}}}                                     \\ \cline{3-8} 
\multicolumn{1}{|c|}{}                               &                                                                                        & \multicolumn{1}{c|}{$\pmb{\gamma = 5}$}           & \multicolumn{1}{c|}{$\pmb{\gamma = 50}$}         & \multicolumn{1}{c|}{$\pmb{\gamma = 100}$} & \multicolumn{1}{c|}{$\pmb{\gamma = 5}$}        & \multicolumn{1}{c|}{$\pmb{\gamma = 50}$}       & \multicolumn{1}{c|}{$\pmb{\gamma = 100}$} \\ \hline
\multicolumn{1}{|c|}{\textbf{BBCA}}                  & 8825                                                                                   & \multicolumn{1}{l|}{Rp0.00}      & \multicolumn{1}{l|}{Rp0.00}     & Rp0.00                   & \multicolumn{1}{l|}{Rp0.00}   & \multicolumn{1}{l|}{Rp0.00}   & Rp0.00                   \\ \hline
\multicolumn{1}{|c|}{\textbf{BBMD}}                  & 1960                                                                                   & \multicolumn{1}{l|}{Rp0.00}      & \multicolumn{1}{l|}{Rp0.00}     & Rp0.00                   & \multicolumn{1}{l|}{Rp0.00}   & \multicolumn{1}{l|}{Rp0.00}   & Rp0.00                   \\ \hline
\multicolumn{1}{|c|}{\textbf{BBNI}}                  & 9350                                                                                   & \multicolumn{1}{l|}{Rp0.00}      & \multicolumn{1}{l|}{Rp0.00}     & Rp0.00                   & \multicolumn{1}{l|}{Rp0.00}   & \multicolumn{1}{l|}{Rp0.00}   & Rp0.00                   \\ \hline
\multicolumn{1}{|c|}{\textbf{BDMN}}                  & 2890                                                                                   & \multicolumn{1}{l|}{Rp0.00}      & \multicolumn{1}{l|}{Rp0.00}     & Rp0.00                   & \multicolumn{1}{l|}{Rp0.00}   & \multicolumn{1}{l|}{Rp0.00}   & Rp0.00                   \\ \hline
\multicolumn{1}{|c|}{\textbf{BINA}}                  & 3990                                                                                   & \multicolumn{1}{l|}{Rp0.00}      & \multicolumn{1}{l|}{Rp0.00}     & Rp0.00                   & \multicolumn{1}{l|}{Rp0.00}   & \multicolumn{1}{l|}{Rp0.00}   & Rp0.00                   \\ \hline
\multicolumn{1}{|c|}{\textbf{BJBR}}                  & 1370                                                                                   & \multicolumn{1}{l|}{Rp0.00}      & \multicolumn{1}{l|}{Rp280.00}   & Rp280.00                 & \multicolumn{1}{l|}{Rp0.00}   & \multicolumn{1}{l|}{Rp350.00} & Rp385.00                 \\ \hline
\multicolumn{1}{|c|}{\textbf{BJTM}}                  & 735                                                                                    & \multicolumn{1}{l|}{Rp0.00}      & \multicolumn{1}{l|}{Rp0.00}     & Rp0.00                   & \multicolumn{1}{l|}{Rp0.00}   & \multicolumn{1}{l|}{Rp0.00}   & Rp0.00                   \\ \hline
\multicolumn{1}{|c|}{\textbf{BMAS}}                  & 1415                                                                                   & \multicolumn{1}{l|}{Rp0.00}      & \multicolumn{1}{l|}{Rp0.00}     & Rp0.00                   & \multicolumn{1}{l|}{Rp0.00}   & \multicolumn{1}{l|}{Rp0.00}   & Rp0.00                   \\ \hline
\multicolumn{1}{|c|}{\textbf{BMRI}}                  & 10225                                                                                  & \multicolumn{1}{l|}{Rp9,550.00}  & \multicolumn{1}{l|}{Rp0.00}     & Rp0.00                   & \multicolumn{1}{l|}{Rp0.00}   & \multicolumn{1}{l|}{Rp0.00}   & Rp0.00                   \\ \hline
\multicolumn{1}{|c|}{\textbf{BNGA}}                  & 1275                                                                                   & \multicolumn{1}{l|}{Rp2,150.00}  & \multicolumn{1}{l|}{Rp600.00}   & Rp450.00                 & \multicolumn{1}{l|}{Rp0.00}   & \multicolumn{1}{l|}{Rp150.00} & Rp150.00                 \\ \hline
\multicolumn{1}{|c|}{\textbf{BNII}}                  & 230                                                                                    & \multicolumn{1}{l|}{Rp0.00}      & \multicolumn{1}{l|}{Rp0.00}     & Rp32.00                  & \multicolumn{1}{l|}{Rp0.00}   & \multicolumn{1}{l|}{Rp0.00}   & Rp40.00                  \\ \hline
\multicolumn{1}{|c|}{\textbf{BNLI}}                  & 945                                                                                    & \multicolumn{1}{l|}{Rp0.00}      & \multicolumn{1}{l|}{Rp0.00}     & Rp15.00                  & \multicolumn{1}{l|}{Rp0.00}   & \multicolumn{1}{l|}{Rp0.00}   & Rp0.00                   \\ \hline
\multicolumn{1}{|c|}{\textbf{BRIS}}                  & 1640                                                                                   & \multicolumn{1}{l|}{Rp0.00}      & \multicolumn{1}{l|}{Rp0.00}     & Rp0.00                   & \multicolumn{1}{l|}{Rp0.00}   & \multicolumn{1}{l|}{Rp0.00}   & Rp0.00                   \\ \hline
\multicolumn{1}{|c|}{\textbf{BSIM}}                  & 890                                                                                    & \multicolumn{1}{l|}{Rp0.00}      & \multicolumn{1}{l|}{Rp0.00}     & Rp0.00                   & \multicolumn{1}{l|}{Rp0.00}   & \multicolumn{1}{l|}{Rp0.00}   & Rp0.00                   \\ \hline
\multicolumn{1}{|c|}{\textbf{BTPN}}                  & 2470                                                                                   & \multicolumn{1}{l|}{Rp0.00}      & \multicolumn{1}{l|}{Rp0.00}     & -Rp20.00                 & \multicolumn{1}{l|}{Rp0.00}   & \multicolumn{1}{l|}{-Rp40.00} & -Rp40.00                 \\ \hline
\multicolumn{1}{|c|}{\textbf{MASB}}                  & 3420                                                                                   & \multicolumn{1}{l|}{Rp0.00}      & \multicolumn{1}{l|}{Rp0.00}     & Rp0.00                   & \multicolumn{1}{l|}{Rp0.00}   & \multicolumn{1}{l|}{Rp0.00}   & Rp0.00                   \\ \hline
\multicolumn{1}{|c|}{\textbf{MEGA}}                  & 5125                                                                                   & \multicolumn{1}{l|}{Rp0.00}      & \multicolumn{1}{l|}{Rp0.00}     & Rp0.00                   & \multicolumn{1}{l|}{Rp0.00}   & \multicolumn{1}{l|}{Rp0.00}   & Rp0.00                   \\ \hline
\multicolumn{1}{|c|}{\textbf{NISP}}                  & 770                                                                                    & \multicolumn{1}{l|}{Rp15.00}     & \multicolumn{1}{l|}{Rp150.00}   & Rp135.00                 & \multicolumn{1}{l|}{Rp60.00}  & \multicolumn{1}{l|}{Rp165.00} & Rp150.00                 \\ \hline
\multicolumn{1}{|c|}{\textbf{PNBN}}                  & 1385                                                                                   & \multicolumn{1}{l|}{Rp0.00}      & \multicolumn{1}{l|}{Rp0.00}     & Rp0.00                   & \multicolumn{1}{l|}{Rp0.00}   & \multicolumn{1}{l|}{Rp0.00}   & Rp0.00                   \\ \hline
\multicolumn{1}{|c|}{\textbf{PNBS}}                  & 60                                                                                     & \multicolumn{1}{l|}{Rp0.00}      & \multicolumn{1}{l|}{Rp0.00}     & Rp0.00                   & \multicolumn{1}{l|}{Rp0.00}   & \multicolumn{1}{l|}{Rp0.00}   & Rp0.00                   \\ \hline
\multicolumn{1}{|c|}{\textbf{SDRA}}                  & 590                                                                                    & \multicolumn{1}{l|}{Rp175.00}    & \multicolumn{1}{l|}{Rp220.00}   & Rp195.00                 & \multicolumn{1}{l|}{Rp450.00} & \multicolumn{1}{l|}{Rp225.00} & Rp200.00                 \\ \hline
\multicolumn{2}{|c|}{\textbf{Total}}                                                                                                          & \multicolumn{1}{l|}{Rp11,715.00} & \multicolumn{1}{l|}{Rp1,030.00} & Rp892.00                 & \multicolumn{1}{l|}{Rp60.00}  & \multicolumn{1}{l|}{Rp625.00} & Rp685.00                 \\ \hline
\end{tabular}
}
\end{table}

Now, we present a different case in which the profit or loss is obtained under poor market conditions, as shown in Table \ref{tab:pasarjelek}. The moving-window method yields profits of Rp. 735.00 for $\gamma = 5$, losses of Rp. 2,040.00 for $\gamma = 50$, and losses of Rp. 2,629.00 for $\gamma = 100$. On the other hand, the bootstrapping method results in losses across all cases Rp. 3,830.00 for $\gamma = 5$, Rp. 3,870.00 for $\gamma = 20$, and Rp. 3,830.00 for $\gamma = 100$.

\begin{table}
\caption{\textit{Capital Gain (Loss)} on Bad Market Condition.}
\label{tab:pasarjelek}
\resizebox{\textwidth}{!}{
\begin{tabular}{|cc|lll|lll|}
\hline
\multicolumn{1}{|c|}{\multirow{2}{*}{\textbf{Asset}}} & \multirow{2}{*}{\textbf{\begin{tabular}[c]{@{}c@{}}Proice\\      (2/5/2023)\end{tabular}}} & \multicolumn{3}{c|}{\textit{\textbf{Moving   Window}}}                                       & \multicolumn{3}{c|}{\textit{\textbf{Bootstrapping}}}                                           \\ \cline{3-8} 
\multicolumn{1}{|c|}{}                               &                                                                                           & \multicolumn{1}{c|}{$\pmb{\gamma = 5}$}         & \multicolumn{1}{c|}{$\pmb{\gamma = 50}$}          & \multicolumn{1}{c|}{$\pmb{\gamma = 100}$} & \multicolumn{1}{c|}{$\pmb{\gamma = 5}$}           & \multicolumn{1}{c|}{$\pmb{\gamma = 50}$}          & \multicolumn{1}{c|}{$\pmb{\gamma = 100}$} \\ \hline
\multicolumn{1}{|c|}{\textbf{BBCA}}                  & 9050                                                                                      & \multicolumn{1}{l|}{Rp225.00}  & \multicolumn{1}{l|}{Rp0.00}      & Rp0.00                   & \multicolumn{1}{l|}{Rp0.00}      & \multicolumn{1}{l|}{Rp0.00}      & Rp0.00                   \\ \hline
\multicolumn{1}{|c|}{\textbf{BBMD}}                  & 1925                                                                                      & \multicolumn{1}{l|}{Rp0.00}    & \multicolumn{1}{l|}{-Rp70.00}    & -Rp105.00                & \multicolumn{1}{l|}{Rp0.00}      & \multicolumn{1}{l|}{Rp0.00}      & -Rp35.00                 \\ \hline
\multicolumn{1}{|c|}{\textbf{BBNI}}                  & 9550                                                                                      & \multicolumn{1}{l|}{Rp0.00}    & \multicolumn{1}{l|}{Rp0.00}      & Rp0.00                   & \multicolumn{1}{l|}{Rp0.00}      & \multicolumn{1}{l|}{Rp0.00}      & Rp0.00                   \\ \hline
\multicolumn{1}{|c|}{\textbf{BDMN}}                  & 2770                                                                                      & \multicolumn{1}{l|}{Rp0.00}    & \multicolumn{1}{l|}{Rp0.00}      & Rp0.00                   & \multicolumn{1}{l|}{Rp0.00}      & \multicolumn{1}{l|}{Rp0.00}      & Rp0.00                   \\ \hline
\multicolumn{1}{|c|}{\textbf{BINA}}                  & 3970                                                                                      & \multicolumn{1}{l|}{-Rp60.00}  & \multicolumn{1}{l|}{-Rp20.00}    & -Rp20.00                 & \multicolumn{1}{l|}{Rp0.00}      & \multicolumn{1}{l|}{Rp0.00}      & Rp0.00                   \\ \hline
\multicolumn{1}{|c|}{\textbf{BJBR}}                  & 1220                                                                                      & \multicolumn{1}{l|}{Rp0.00}    & \multicolumn{1}{l|}{-Rp920.00}   & -Rp920.00                & \multicolumn{1}{l|}{Rp0.00}      & \multicolumn{1}{l|}{-Rp1,150.00} & -Rp1,265.00              \\ \hline
\multicolumn{1}{|c|}{\textbf{BJTM}}                  & 665                                                                                       & \multicolumn{1}{l|}{Rp0.00}    & \multicolumn{1}{l|}{-Rp1,890.00} & -Rp2,310.00              & \multicolumn{1}{l|}{-Rp4,270.00} & \multicolumn{1}{l|}{-Rp3,990.00} & -Rp3,710.00              \\ \hline
\multicolumn{1}{|c|}{\textbf{BMAS}}                  & 1235                                                                                      & \multicolumn{1}{l|}{Rp0.00}    & \multicolumn{1}{l|}{Rp0.00}      & Rp0.00                   & \multicolumn{1}{l|}{Rp0.00}      & \multicolumn{1}{l|}{Rp0.00}      & Rp0.00                   \\ \hline
\multicolumn{1}{|c|}{\textbf{BMRI}}                  & 5250                                                                                      & \multicolumn{1}{l|}{-Rp400.00} & \multicolumn{1}{l|}{Rp0.00}      & Rp0.00                   & \multicolumn{1}{l|}{Rp0.00}      & \multicolumn{1}{l|}{Rp0.00}      & Rp0.00                   \\ \hline
\multicolumn{1}{|c|}{\textbf{BNGA}}                  & 1245                                                                                      & \multicolumn{1}{l|}{Rp860.00}  & \multicolumn{1}{l|}{Rp240.00}    & Rp180.00                 & \multicolumn{1}{l|}{Rp0.00}      & \multicolumn{1}{l|}{Rp60.00}     & Rp60.00                  \\ \hline
\multicolumn{1}{|c|}{\textbf{BNII}}                  & 228                                                                                       & \multicolumn{1}{l|}{Rp0.00}    & \multicolumn{1}{l|}{Rp0.00}      & Rp16.00                  & \multicolumn{1}{l|}{Rp0.00}      & \multicolumn{1}{l|}{Rp0.00}      & Rp20.00                  \\ \hline
\multicolumn{1}{|c|}{\textbf{BNLI}}                  & 950                                                                                       & \multicolumn{1}{l|}{Rp0.00}    & \multicolumn{1}{l|}{Rp0.00}      & Rp20.00                  & \multicolumn{1}{l|}{Rp0.00}      & \multicolumn{1}{l|}{Rp0.00}      & Rp0.00                   \\ \hline
\multicolumn{1}{|c|}{\textbf{BRIS}}                  & 1685                                                                                      & \multicolumn{1}{l|}{Rp0.00}    & \multicolumn{1}{l|}{Rp0.00}      & Rp0.00                   & \multicolumn{1}{l|}{Rp0.00}      & \multicolumn{1}{l|}{Rp0.00}      & Rp0.00                   \\ \hline
\multicolumn{1}{|c|}{\textbf{BSIM}}                  & 890                                                                                       & \multicolumn{1}{l|}{Rp0.00}    & \multicolumn{1}{l|}{Rp0.00}      & Rp0.00                   & \multicolumn{1}{l|}{Rp0.00}      & \multicolumn{1}{l|}{Rp0.00}      & Rp0.00                   \\ \hline
\multicolumn{1}{|c|}{\textbf{BTPN}}                  & 2490                                                                                      & \multicolumn{1}{l|}{Rp0.00}    & \multicolumn{1}{l|}{Rp0.00}      & Rp0.00                   & \multicolumn{1}{l|}{Rp0.00}      & \multicolumn{1}{l|}{Rp0.00}      & Rp0.00                   \\ \hline
\multicolumn{1}{|c|}{\textbf{MASB}}                  & 3180                                                                                      & \multicolumn{1}{l|}{Rp0.00}    & \multicolumn{1}{l|}{-Rp480.00}   & -Rp480.00                & \multicolumn{1}{l|}{Rp0.00}      & \multicolumn{1}{l|}{Rp0.00}      & Rp0.00                   \\ \hline
\multicolumn{1}{|c|}{\textbf{MEGA}}                  & 4980                                                                                      & \multicolumn{1}{l|}{Rp0.00}    & \multicolumn{1}{l|}{Rp0.00}      & Rp0.00                   & \multicolumn{1}{l|}{Rp0.00}      & \multicolumn{1}{l|}{Rp0.00}      & Rp0.00                   \\ \hline
\multicolumn{1}{|c|}{\textbf{NISP}}                  & 865                                                                                       & \multicolumn{1}{l|}{Rp110.00}  & \multicolumn{1}{l|}{Rp1,100.00}  & Rp990.00                 & \multicolumn{1}{l|}{Rp440.00}    & \multicolumn{1}{l|}{Rp1,210.00}  & Rp1,100.00               \\ \hline
\multicolumn{1}{|c|}{\textbf{PNBN}}                  & 1040                                                                                      & \multicolumn{1}{l|}{Rp0.00}    & \multicolumn{1}{l|}{Rp0.00}      & Rp0.00                   & \multicolumn{1}{l|}{Rp0.00}      & \multicolumn{1}{l|}{Rp0.00}      & Rp0.00                   \\ \hline
\multicolumn{1}{|c|}{\textbf{PNBS}}                  & 58                                                                                        & \multicolumn{1}{l|}{Rp0.00}    & \multicolumn{1}{l|}{Rp0.00}      & Rp0.00                   & \multicolumn{1}{l|}{Rp0.00}      & \multicolumn{1}{l|}{Rp0.00}      & Rp0.00                   \\ \hline
\multicolumn{1}{|c|}{\textbf{SDRA}}                  & 560                                                                                       & \multicolumn{1}{l|}{-Rp875.00} & \multicolumn{1}{l|}{-Rp1,100.00} & -Rp975.00                & \multicolumn{1}{l|}{-Rp2,250.00} & \multicolumn{1}{l|}{-Rp1,125.00} & -Rp1,000.00              \\ \hline
\multicolumn{2}{|c|}{\textbf{Total}}                                                                                                             & \multicolumn{1}{l|}{Rp735.00}  & \multicolumn{1}{l|}{-Rp2,040.00} & -Rp2,629.00              & \multicolumn{1}{l|}{-Rp3,830.00} & \multicolumn{1}{l|}{-Rp3,870.00} & -Rp3,830.00              \\ \hline
\end{tabular}
}
\end{table}

Based on these two market conditions and the three values of risk aversion, it can be observed that the portfolio generated by robust optimization using the moving-window method yields profits in both good and bad market conditions, particularly when $\gamma = 5$.. This indicates that robust optimization with the moving-window method produces a resilient portfolio across different market scenarios, especially for risk-taking investors.

When assessing stock market performance under both good and poor market conditions, price movements from 27 March 2023 to 15 June 2023 suggest that the market can be classified as in good condition if the overall increase in stock prices is significant. However, identifying market conditions solely from short-term price fluctuations remains challenging.

\begin{figure}[h!]
 \centering
 \includegraphics[width=\textwidth]{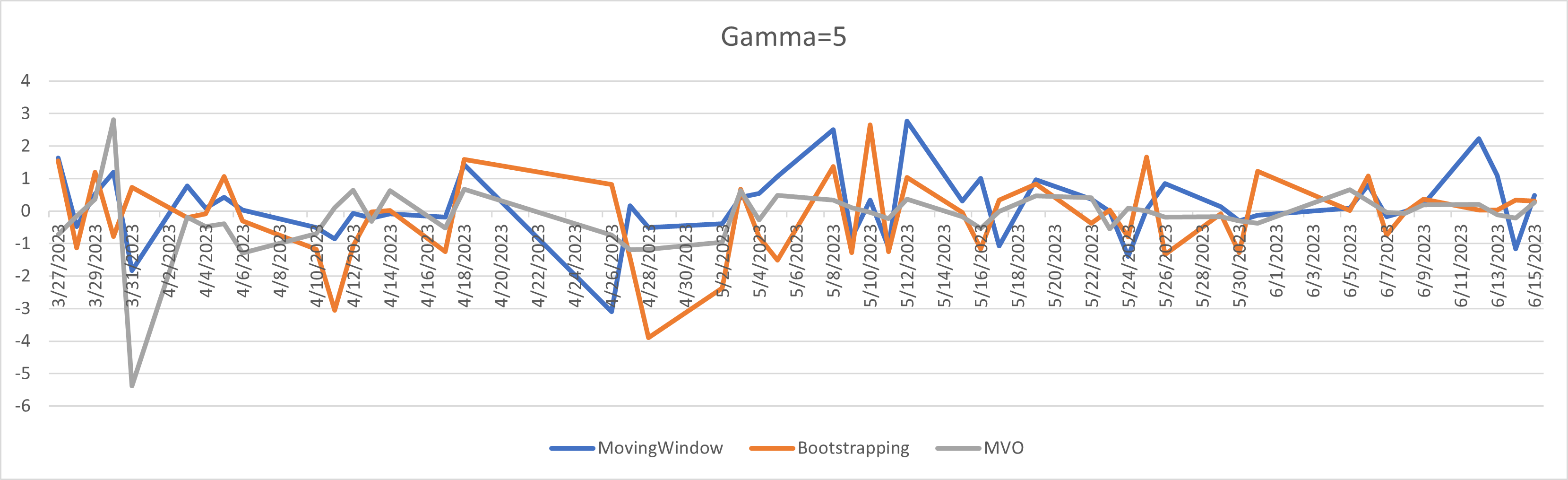}
	\caption{Portfolio Return from 27 March 2023 to 15 June 2023 for $\gamma = 5$.}
	\label{PergerekanReturn5}
\end{figure}

\begin{figure}[h!]
 \centering
 \includegraphics[width=\textwidth]{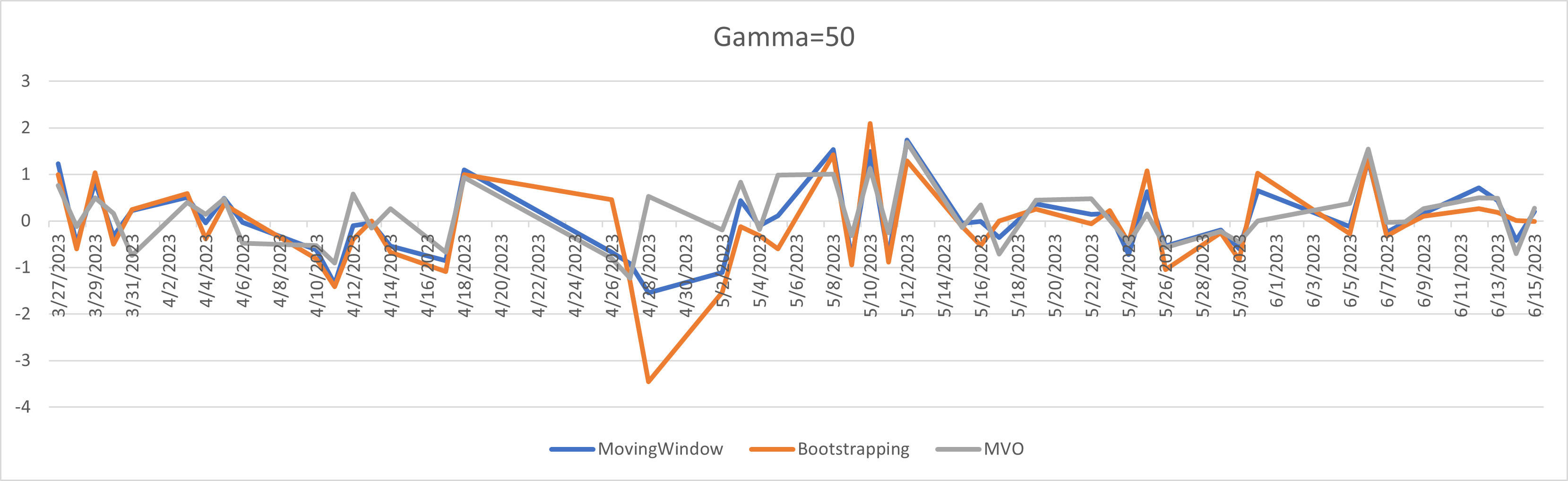}
	\caption{Portfolio Return from 27 March 2023 to 15 June 2023 for $\gamma = 50$.}
	\label{PergerekanReturn50}
\end{figure}

\begin{figure}[h!]
 \centering
 \includegraphics[width=\textwidth]{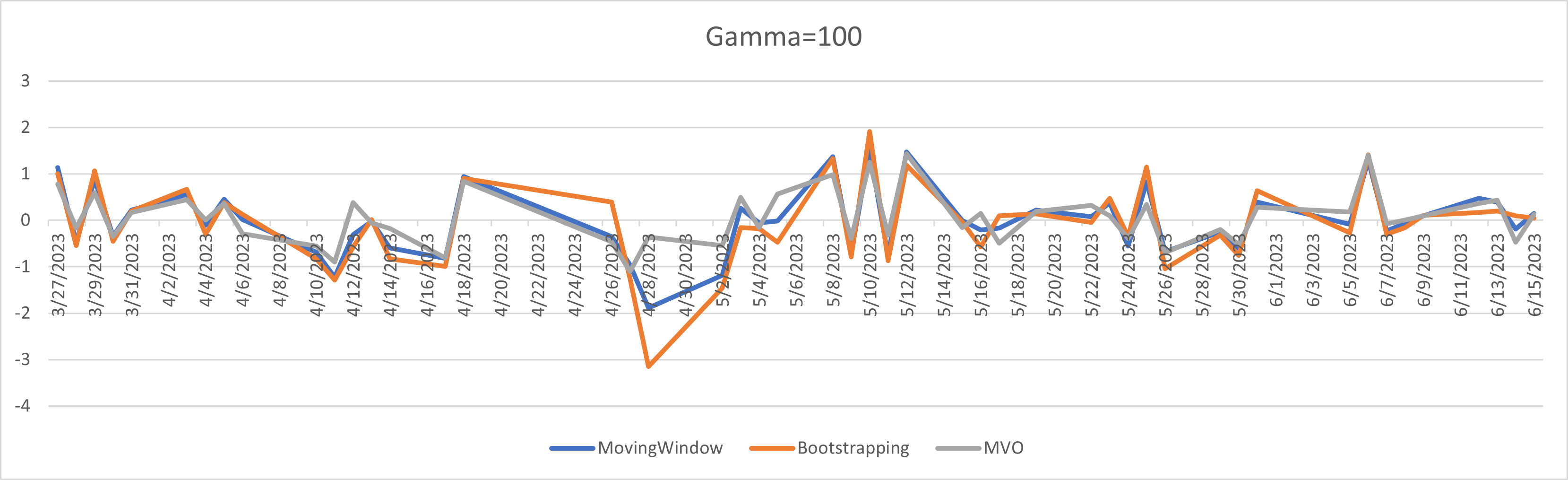}
	\caption{Portfolio Return from 27 March 2023 to 15 June 2023 for $\gamma = 100$.}
	\label{PergerekanReturn100}
\end{figure}

A comparison of portfolio returns is conducted for the 19 selected stocks in Table \ref{tab:proporsi}, which are expected to be chosen by investors using the three optimization methods. The portfolio return dynamics over the period from 27 March 2023 to 15 June 2023 are shown in Figure \ref{PergerekanReturn5}, Figure \ref{PergerekanReturn50}, and Figure \ref{PergerekanReturn100}. For $\gamma = 5$, robust optimization with the moving-window method produced higher profits, as its return curve is consistently positioned above those of the other optimization methods. For higher values of $\gamma$, namely 50 and 100, the results show that mean-variance optimization yielded greater profits, with its return curve predominantly above the others. This demonstrates that for lower $\gamma$ values—corresponding to more risk-taking investors—robust optimization with the moving-window method is preferable to achieve higher profits. Conversely, for risk-averse investors (higher $\gamma$ values), mean-variance optimization is more suitable, as it provides higher profits.

\section{Conclusion}
Based on the conducted experiments, both the moving-window method and the bootstrapping method were employed to construct uncertainty sets for the mean and covariance. These uncertainty sets were then used as inputs for robust optimization, influencing the resulting asset allocations within the portfolio to maximize profits. The procedure for determining uncertainty sets was applied consistently across assets, yielding 45 uncertainty intervals for both the mean and covariance.

The results demonstrate that robust optimization with uncertainty sets derived from the moving-window method offers better performance for investors. First, across different $\gamma$ values, the objective function values obtained from the moving-window method are consistently lower than those from the bootstrapping method, indicating a more favorable risk–return trade-off. Second, under both good and poor market conditions, robust optimization with the moving-window method generated profits, suggesting that investors can achieve positive returns regardless of market fluctuations. Finally, portfolio return graphs across different $\gamma$ values show that the moving-window method delivers higher profits when $\gamma = 5$, corresponding to risk-taking investors, while still maintaining competitive performance under higher $\gamma$ values.

\section*{Acknowledgment}
This research was supported by the Departmental Research Funding Program, Batch 1 (2025), under Contract No. 2320/PKS/ITS/2025, between the Directorate of Research and Community Service, Institut Teknologi Sepuluh Nopember (ITS), and Sena Safarina, S.Si., M.Sc., D.Sc.

\bibliographystyle{bibtex/splncs03} %or splncs03_unsrt
\bibliography{references}

%
% ---- Bibliography ----
%
%\begin{thebibliography}{6}
%%
%
%\bibitem {smit:wat}
%Smith, T.F., Waterman, M.S.: Identification of common molecular subsequences.
%J. Mol. Biol. 147, 195?197 (1981). \url{doi:10.1016/0022-2836(81)90087-5}
%
%\bibitem {may:ehr:stein}
%May, P., Ehrlich, H.-C., Steinke, T.: ZIB structure prediction pipeline:
%composing a complex biological workflow through web services.
%In: Nagel, W.E., Walter, W.V., Lehner, W. (eds.) Euro-Par 2006.
%LNCS, vol. 4128, pp. 1148?1158. Springer, Heidelberg (2006).
%\url{doi:10.1007/11823285_121}
%
%\bibitem {fost:kes}
%Foster, I., Kesselman, C.: The Grid: Blueprint for a New Computing Infrastructure.
%Morgan Kaufmann, San Francisco (1999)
%
%\bibitem {czaj:fitz}
%Czajkowski, K., Fitzgerald, S., Foster, I., Kesselman, C.: Grid information services
%for distributed resource sharing. In: 10th IEEE International Symposium
%on High Performance Distributed Computing, pp. 181?184. IEEE Press, New York (2001).
%\url{doi: 10.1109/HPDC.2001.945188}
%
%\bibitem {fo:kes:nic:tue}
%Foster, I., Kesselman, C., Nick, J., Tuecke, S.: The physiology of the grid: an open grid services architecture for distributed systems integration. Technical report, Global Grid
%Forum (2002)
%
%\bibitem {onlyurl}
%National Center for Biotechnology Information. \url{http://www.ncbi.nlm.nih.gov}
%
%
%\end{thebibliography}
\end{document}